# A Short Image Series Based Scheme for Time Series Digital Image Correlation

Xian Wang, Shaopeng Ma*

*Key Laboratory of Dynamics and Control of Flight Vehicle, Ministry of Education, School of Aerospace Engineering, Beijing Institute of Technology, Beijing 100081, China*

## Abstract

A new scheme for digital image correlation (DIC), i.e., short time series DIC (STS-DIC) is proposed. Instead of processing the original deformed speckle images individually, STS-DIC combines several adjacent deformed speckle images from a short time series and then processes the "averaged image", for which deformation continuity over time is introduced. The deformation of several adjacent images is assumed to be linear in time and a new spatial-temporal displacement representation method with eight unknowns is presented based on the subset-based representation method. Then, the model of STS-DIC is created and a solving scheme is developed based on the Newton-Raphson iteration. The proposed method is verified for numerical and experimental cases. The results show that the proposed STS-DIC greatly improves the accuracy of traditional DIC, both under simple and complicated deformation conditions, while retaining acceptable actual computational cost.

## Introduction

Digital image correlation (DIC) [1, 2] has been widely used in various fields, because of its ability to realize whole field deformation measurement and its convenience in terms of experimental preparation and data processing. To implement the deformation measurement using DIC, the surface of the specimen is first marked with speckle patterns (natural texture or artificial random patterns). Then, one speckle image, called the reference image, is captured before loading and taken as the reference for DIC. Thereafter, the specimen is loaded and the speckle images, called the deformed images, are captured during the loading process. Finally, to obtain the deformation fields at different times, the corresponding deformed images are "compared" individually with the reference image.

Improvement in the measurement accuracy of DIC is always a hot issue and many studies have been introduced in recent years. Besides the error from hardware [3] or from the experimental setup [4, 5], the error from DIC algorithms has also been systematically studied and several improvements have been suggested, including the expressions of correlation coefficients [6] and image interpolation methods [7] in the traditional DIC scheme. Meanwhile, various new DIC schemes have also been developed by introducing a new mechanical mechanism in DIC. For example, Sun [8], Réthoré [9], and Ma [10], amongst others, introduced spatial continuity of displacement into DIC and developed several global DIC schemes, with improved accuracy of DIC on measuring the heterogeneous field. Similar to spatial continuity, the displacement to be measured by DIC is also

*Corresponding author. *Tel.*: +86 10 68912321. *E-mail*: masp@bit.edu.cn.

continuous over time in most circumstances. Currently, owing to cheap and convenient image recording equipment, i.e., digital CCD or CMOS cameras, generally hundreds or thousands of speckle images are captured during the loading procedure for one experiment. Thus, the deformation of an image series can be continuous even in a fast process. However, temporal continuity of deformation is not considered in the traditional DIC scheme. As described above, the deformed images are analyzed independently with respect to time. Actually, in most DIC analyses, only a small percentage of the deformed images are analyzed with most of the images totally wasted.

If the time series information is used, i.e., temporal continuity of displacement is introduced into the DIC scheme as an extra constraint, the accuracy of DIC is expected to improve. Besnard et al. [11, 12] first noticed this and proposed a space-time approach for DIC. In their approach, all the captured images are combined into a "volume" by adding time as the third dimension (the approach is called movie-DIC in this paper). Then, the displacement shape function on the space-time dimensions, instead of only on the space dimension as in traditional DIC, is used and the deformation is solved in the image volume. As a result, a spatial-temporal deformation, i.e., deformation field evolution, can be obtained. Using this approach, deformation continuity over time is introduced into the DIC scheme, with their examples showing the improvement in accuracy of DIC. Owing to the introduction of speckle images on the time series, this new scheme could also be called time series DIC (abbreviated as TS-DIC). Moreover, owing to the introduction of the complete time series, the computational requirements of movie-DIC are remarkably increased compared with that of traditional DIC, similar to the comparison of computation cost between VIC [13] and traditional DIC. This greatly affects the practical use of the approach. Meanwhile, the flexibility of the approach can be improved in practical use. In a DIC experiment, it is common to analyze a specific image (or images) in a very short time. However, this is still very time consuming for movie-DIC because all the images (at least the images before the processed image) must be processed. Moreover, the final accuracy of movie-DIC may be affected by the complexity of the optimization and the mismatch of the deformation shape function. The degrees of freedom, i.e., the number of unknowns for optimizations of movie-DIC are much larger than those for traditional-DIC, even much larger than for FEM-DIC, while the numerical errors from optimization are remarkable. The deformation over time could be very complicated, for example, a homogeneous deformation followed by a crack initialization and propagation, such that the displacement representation with unified shape functions may mismatch the real deformation, thereby introducing additional errors.

In this paper, a new scheme for TS-DIC is proposed. The scheme is merely a modified version of the traditional subset-based DIC scheme, with the addition of a simple assumption to the model. The scheme is constructed to process the reference image and a short time series (STS) in a complete image series, and is called STS-DIC. In STS-DIC, the deformation of several adjacent images is assumed to be linear in time and these are then treated as one special "averaged" image. Then, the DIC model is constructed between the reference image and the "averaged" deformed image. A solving scheme for STS-DIC is also developed based on traditional DIC, merely by adding two more unknowns to the optimization. To process the complete image series, the continuous sectional approach is applied, by which continuity over time is guaranteed. Because of the processing of the short image series and the sectional approach, STS-DIC has very low time consumption and high flexibility in practical use. The time consumption is tested as 1~2 orders higher than traditional DIC, which is acceptable. Processing of specific images with complicated deformation and without the "support" of previous images is also possible in STS-DIC. Therefore, it is expected that the developed STS-DIC would be more applicable in practical use.

The paper is organized as follows: In the next section, the STS-DIC model is constructed and the solving scheme is developed. In the third section, the proposed method is verified for numerical and experimental cases with simple and complicated deformations. Finally, the paper concludes in the last section.

## Principle of STS-DIC

Let $f$ represent the reference speckle image, and $g_t$ ($t = t_1, t_2, \ldots, t_n$) represent the series of deformed speckle images captured at time $t$. According to the "constant intensity assumption", there exists an equation for each of the image pairs, $f$ and $g_t$, such that

$$f\left(\mathbf{X}\right)=g_t\left(\mathbf{X}+\mathbf{U}\left(\mathbf{X},t\right)\right) \qquad t=t_1,t_2,\cdots,t_n \,, \tag{1}$$

where $\mathbf{X}$ is the coordinates of the pixels, and $\mathbf{U}(\mathbf{X}, t)$ is the displacement of image $g_t$ referring to $f$.

For the $2m$+1 ($m$ = 1, 2, 3, …) deformed speckle images captured between [$t_{i-m}$ $t_{i+m}$], a combined equation can be written as

$$f\left(\mathbf{X}\right)=\frac{1}{2m+1}\int_{t_{i-m}}^{t_{i+m}} g_t\left(\mathbf{X}+\mathbf{U}\left(\mathbf{X},t\right)\right)dt \,. \tag{2}$$

The right-hand side of Eq. 2, the special "average" of $2m$+1 images, can be regarded as a new "speckle image", denoted by

$$G_i^{(m)}\left(\mathbf{X}+\mathbf{U}\left(\mathbf{X},t\right)\right)=\frac{1}{2m+1}\int_{t_{i-m}}^{t_{i+m}} g_t\left(\mathbf{X}+\mathbf{U}\left(\mathbf{X},t\right)\right)dt \,, \tag{3}$$

for simplicity.

The "average" image shown in Eq. 3 could be expected to improve the quality of the speckle images under certain circumstances. For a particular case, when $\mathbf{U}(\mathbf{X}, t) = 0$, i.e., there is no displacement between the $2m$+1 images, $G_i^{(m)}\left(\mathbf{X}+0\right)$ denotes the direct average of the "same" $2m$+1 digital images captured at different times. Based on probability theory, noise in the new averaged image decreases $\sqrt{2m+1}$ times compared with the original single speckle image $g_{t_i}$ [14]. For images with non-zero displacement, i.e., $\mathbf{U}(\mathbf{X}, t) \neq 0$, it is still expected that the average of the compensated speckle images, i.e., $g_t\left(\mathbf{X}+\mathbf{U}\left(\mathbf{X},t\right)\right)$, has the same effect as direct superposing of the "same" speckle images. That is, $G_i^{(m)}\left(\mathbf{X}+\mathbf{U}\left(\mathbf{X},t\right)\right)$ is a high quality "speckle image". Therefore, better results can be obtained if a new DIC scheme were developed based on Eq. 2.

Equation 2 is actually a new expression of the "constant intensity assumption" between one reference image and a short series of deformed images. A new DIC scheme could be developed from the equation, as is done in traditional DIC schemes. To solve Eq. 2, the dimensions of the unknowns must be reduced to avoid ill conditions of the equation. Thus, displacement **U**(**X**, *t*) should be approximately represented by certain special functions, as in traditional DIC. Different from traditional DIC, the **U**(**X**, *t*) here should be represented in both the temporal and spatial spaces.

First, we consider the temporal space. Suppose that the deformed speckle images are captured continuously with short time intervals during the experiment. Then, **U**(**X**, *t*) is expected to be a continuous function over time. This is reasonable when considering deformation continuity over time on a solid specimen. Therefore, **U**(**X**, *t*) can be expanded with *t* by Taylor expansion as

$$\mathbf{U}(\mathbf{X},t)=\mathbf{U}(\mathbf{X},t)\big|_{t=t_i}+\frac{\partial\mathbf{U}(\mathbf{X},t)}{\partial t}\bigg|_{t=t_i}\Delta t+\mathrm{o}\left(\Delta t^2\right), \tag{4}$$

where $\Delta t = t - t_i$ is the time interval.

Omitting the higher order small quantity, **U**(**X**, *t*) can be expressed as

$$\mathbf{U}(\mathbf{X},t)\approx\mathbf{U}(\mathbf{X},t)\big|_{t=t_i}+\frac{\partial\mathbf{U}(\mathbf{X},t)}{\partial t}\bigg|_{t=t_i}\Delta t\ . \tag{5}$$

Equation 5 shows that the displacement of image *t* can be represented as the displacement of image $t_i$ plus the multiplication of the velocity of $t_i$ and the time interval between them. As shown in Fig. 1, if the images are captured continuously with a short time interval and the deformation is slow, the displacement of a point within a short period of time can actually be expressed by a linear expression. Therefore, the representation in Eq. 5 is reasonable.

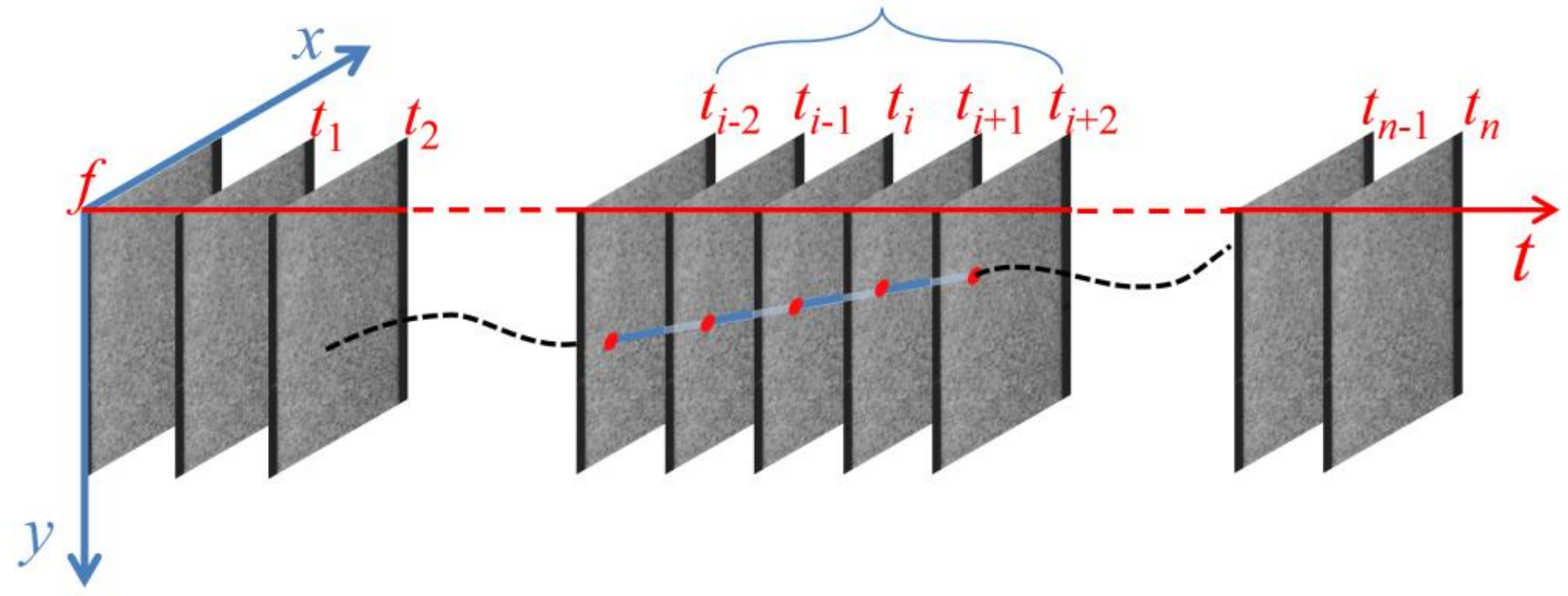

**Fig. 1** Schematic of STS-DIC

Next, we consider dimension reduction in spatial space. The shape function method used in traditional DIC is also used here. Based on the space shape function **Φ**(**X**), the displacement field of image $t_i$ is represented as

$$\mathbf{U}(\mathbf{X},t)\big|_{t=t_i}=\mathbf{\Phi}(\mathbf{X})\mathbf{b}\big|_{t=t_i}\ , \tag{6}$$

where **b** denotes the parameter matrix.

Figure 2 shows one solution for this representation, a type of mesh-based representation. In this case, the displacement interpolation function of the 4-node-quadralateral in FEM acts as the shape function while the displacement of all the nodes in the mesh acts as the parameter matrix [8]. Besides this, there are also various other representations [9, 10, 15 and 16]. Generally, the dimensions of **b** are much lower and Eq. 2 can be solved.

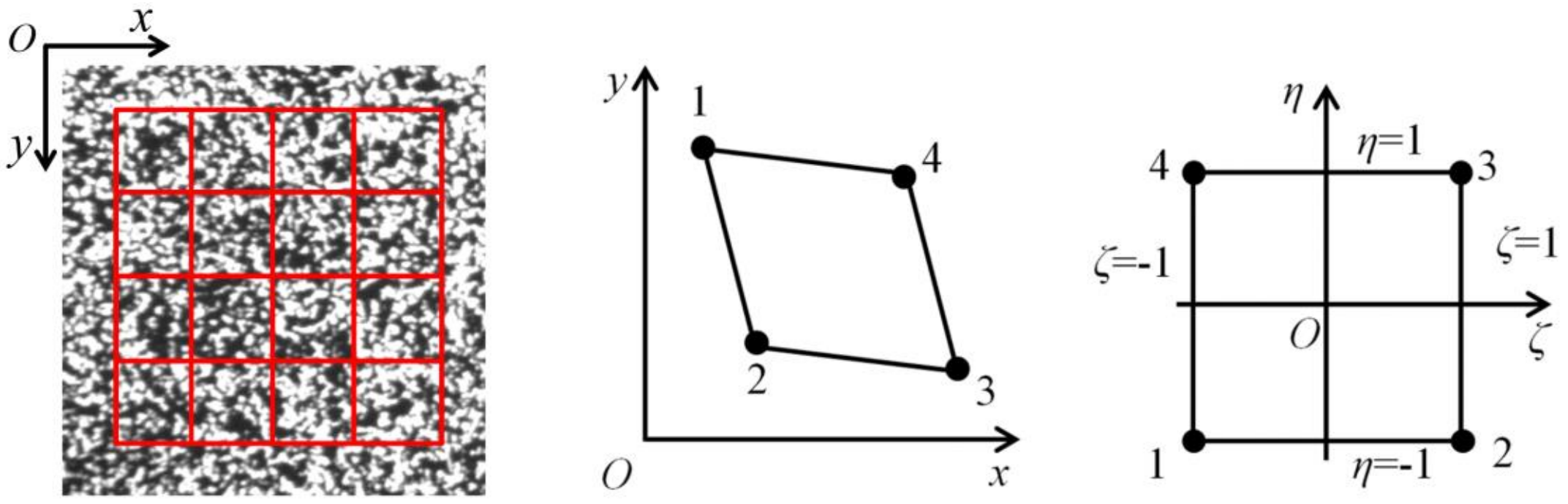

**(a)** **(b)** **(c)**

**Fig. 2** Displacement field representation using 4-node-quadralateral elements: **(a)** the mesh, **(b)** the overall coordinates, and **(c)** the local coordinates of one element

The solving procedure can be further simplified if spatial continuity of deformation is not considered. As is done in subset-based DIC, Eq. 2 can be solved separately on several separate subsets (as shown in Fig. 3a), and the displacement field is finally obtained by combining the results of the different subsets.

Based on the shape function used in traditional subset DIC (as shown in Fig. 3b), the displacement $\mathbf{U}(\mathbf{X},t)\big|_{t=t_i}$ of point Q coordinated at **X** can be expressed as

$$\mathbf{U}(\mathbf{X},t)\big|_{t=t_i} = \begin{bmatrix} u \\ v \end{bmatrix}\Bigg|_{\substack{t=t_i \\ \mathbf{X}=\mathbf{X}_c}} + \begin{bmatrix} \frac{\partial u}{\partial x} & \frac{\partial u}{\partial y} \\ \frac{\partial v}{\partial x} & \frac{\partial v}{\partial y} \end{bmatrix}\Bigg|_{\substack{t=t_i \\ \mathbf{X}=\mathbf{X}_c}} \begin{bmatrix} \Delta x \\ \Delta y \end{bmatrix}, \tag{7}$$

where $\mathbf{X}_c$, which is coordinated at $(x_0\ y_0)$, is the center point of the subset, $x$, $y$ are the horizontal and vertical components of coordinate **X**, $u$, $v$ are the horizontal and vertical components of displacement $\mathbf{U}(\mathbf{X},t)\big|_{t=t_i}$, and $\Delta x$ and $\Delta y$ are the coordinate differences between $P$ and $Q$.

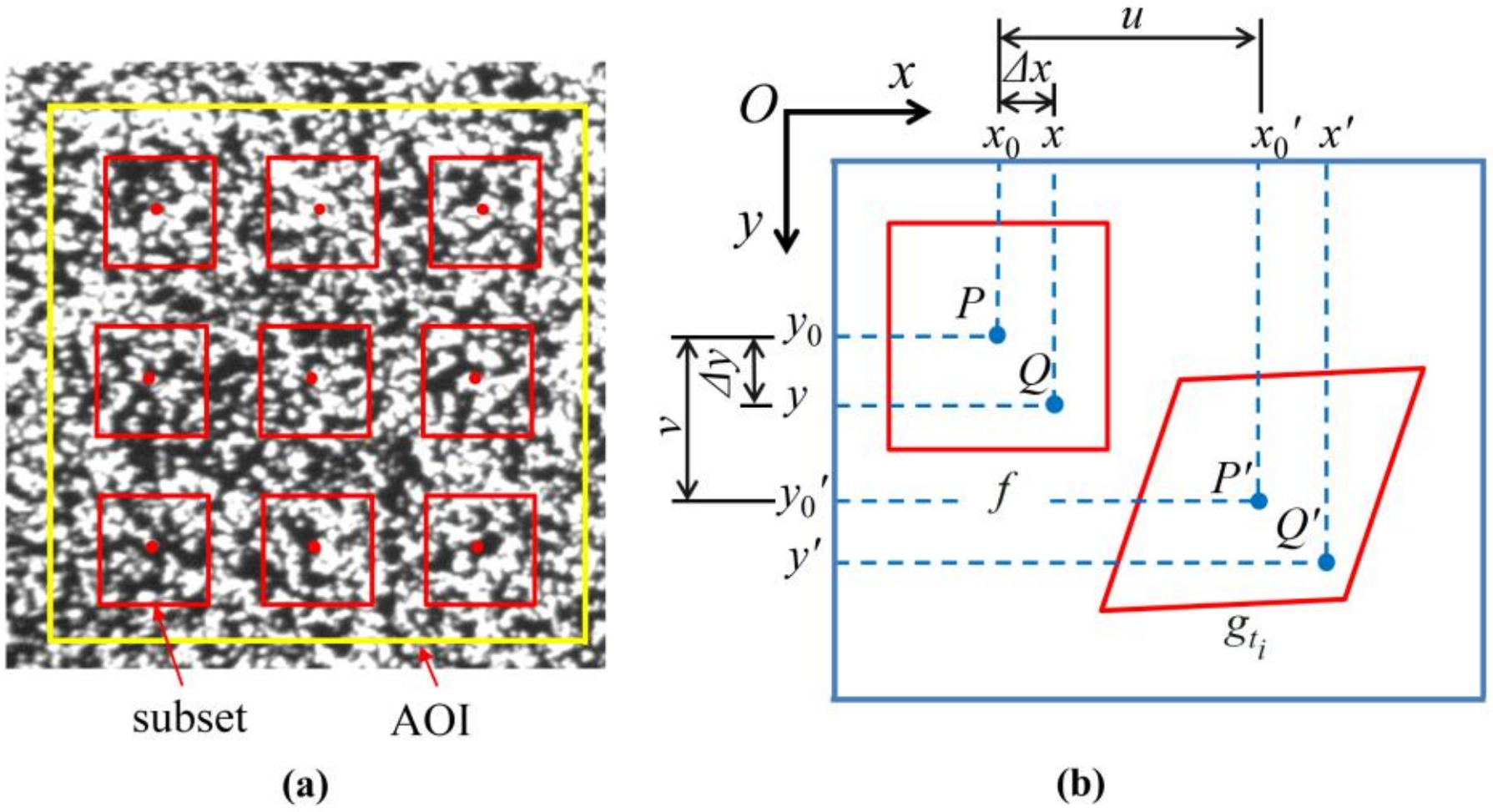

**Fig. 3** Displacement field representation using rectangular subsets: **(a)** partitioning of the area of interest (AOI), and **(b)** the displacement representation of one subset

The velocity item in Eq. 5, $\left.\frac{\partial \mathbf{U}(\mathbf{X},t)}{\partial t}\right|_{t=t_i}$, can then be calculated as

$$\left.\frac{\partial \mathbf{U}(\mathbf{X},t)}{\partial t}\right|_{t=t_i} = \left.\begin{bmatrix} \frac{\partial u}{\partial t} \\ \frac{\partial v}{\partial t} \end{bmatrix}\right|_{\substack{t=t_i \\ \mathbf{x}=\mathbf{x}_c}} + \left.\begin{bmatrix} \frac{\partial^2 u}{\partial x \partial t} & \frac{\partial^2 u}{\partial y \partial t} \\ \frac{\partial^2 v}{\partial x \partial t} & \frac{\partial^2 v}{\partial y \partial t} \end{bmatrix}\right|_{\substack{t=t_i \\ \mathbf{x}=\mathbf{x}_c}} \begin{bmatrix} \Delta x \\ \Delta y \end{bmatrix}. \tag{8}$$

Substituting Eqs. 7 and 8 into Eq. 5, and ignoring the higher order terms $\Delta x\Delta t$ and $\Delta y\Delta t$, the displacement of point Q($x$, $y$) at $t$ (shown in Fig. 4) can be expressed as

$$\mathbf{U}(\mathbf{X},t) = \mathbf{U}(\mathbf{X},t;\mathbf{a}) = \left.\begin{bmatrix} u \\ v \end{bmatrix}\right|_{\substack{t=t_i \\ \mathbf{x}=\mathbf{x}_c}} + \left.\begin{bmatrix} \frac{\partial u}{\partial x} & \frac{\partial u}{\partial y} \\ \frac{\partial v}{\partial x} & \frac{\partial v}{\partial y} \end{bmatrix}\right|_{\substack{t=t_i \\ \mathbf{x}=\mathbf{x}_c}} \begin{bmatrix} \Delta x \\ \Delta y \end{bmatrix} + \left.\begin{bmatrix} \frac{\partial u}{\partial t} \\ \frac{\partial v}{\partial t} \end{bmatrix}\right|_{\substack{t=t_i \\ \mathbf{x}=\mathbf{x}_c}} [\Delta t] = \mathbf{a} \begin{bmatrix} 1 \\ \Delta x \\ \Delta y \\ \Delta t \end{bmatrix}, \tag{9}$$

where

$$\mathbf{a} = \left.\begin{bmatrix} u & \frac{\partial u}{\partial x} & \frac{\partial u}{\partial y} & \frac{\partial u}{\partial t} \\ v & \frac{\partial v}{\partial x} & \frac{\partial v}{\partial y} & \frac{\partial v}{\partial t} \end{bmatrix}\right|_{\substack{t=t_i \\ \mathbf{x}=\mathbf{x}_c}} \tag{10}$$

is the parameter matrix to be solved in STS-DIC. Compared with traditional DIC, two more unknowns, $\left.\frac{\partial u}{\partial t}\right|_{\substack{t=t_i \\ \mathbf{x}=\mathbf{x}_c}}$ and $\left.\frac{\partial v}{\partial t}\right|_{\substack{t=t_i \\ \mathbf{x}=\mathbf{x}_c}}$ are added. The relation in Eq. 9 is expressed as shown in Fig. 4. It

is seen that the displacement of point $Q$ can be expressed in three parts, i.e., the rigid body translation from $P$ to $P'$, the linear spatial deformation between $P'$ to $Q'$, and the linear temporal deformation from $Q'$ to $Q''$.

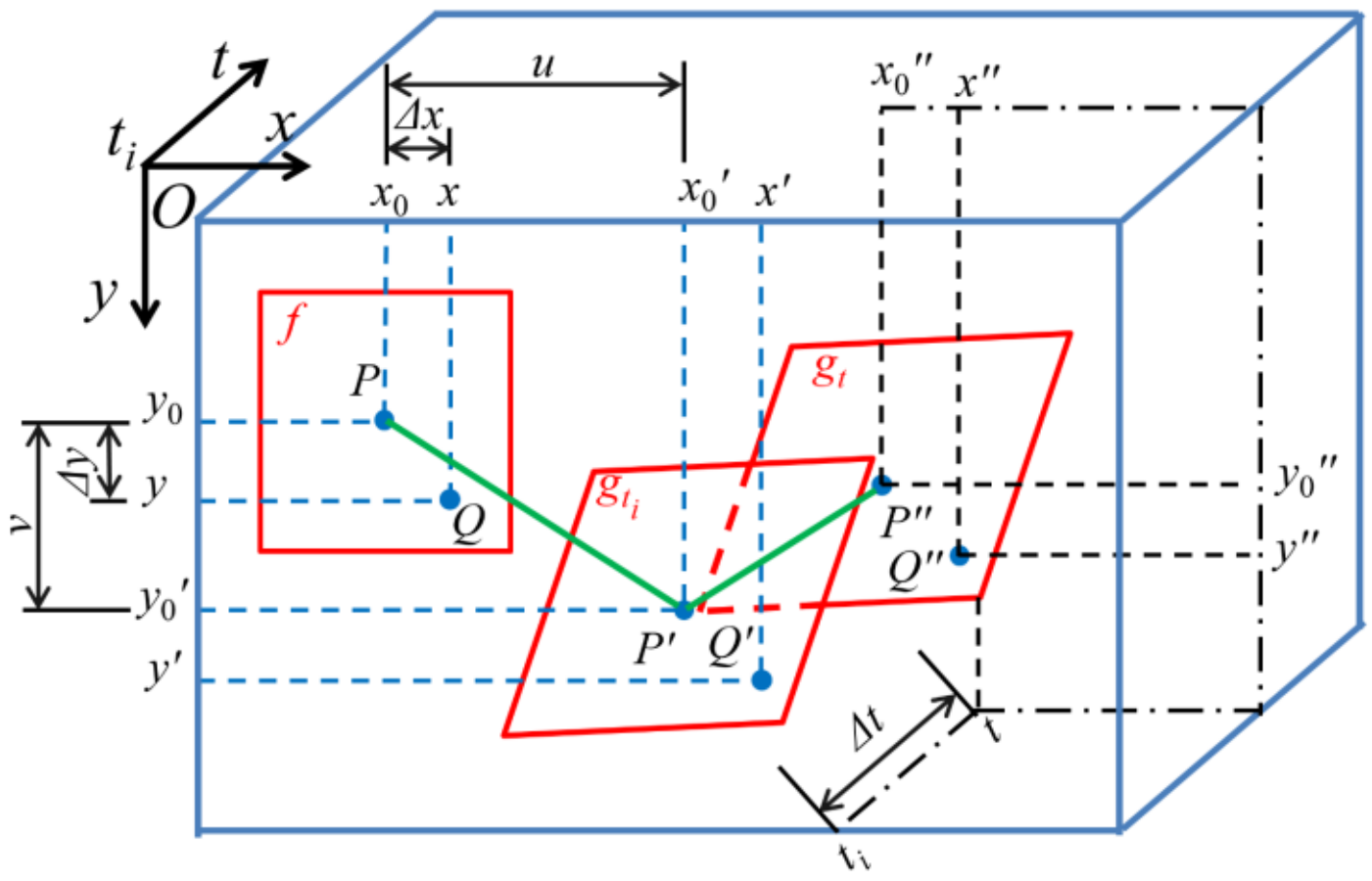

**Fig. 4** Schematic of displacement representation in STS-DIC

Substituting Eq. 9 into Eq. 2, the final expression of the “constant intensity assumption” is obtained, i.e.,

$$f\left(\mathbf{X}\right)=G_i^{(m)}\left(\mathbf{X}+\mathbf{U}\left(\mathbf{X},t;\mathbf{a}\right)\right)=\frac{1}{2m+1}\int_{t_{i-m}}^{t_{i+m}} g_t\left(\mathbf{X}+\mathbf{a}\begin{bmatrix}1 & \Delta x & \Delta y & \Delta t\end{bmatrix}^{\mathrm{T}}\right)dt\ . \tag{11}$$

It is seen that Eq. 11 degrades to the normal Eq. 1 when $m$=0, i.e., only one deformed image is considered. Therefore, traditional DIC is a special case of STS-DIC.

Practically, the number of pixels in the subset (for example 21×21) is much larger than the number of unknowns of **a**, i.e., eight, such that Eq. 11 is over-determined and can be solved using an optimization method such as

$$\min_{\mathbf{a}} C\left(f\left(\mathbf{X}\right),G_i^{(m)}\left(\mathbf{X}+\mathbf{U}\left(\mathbf{X},t;\mathbf{a}\right)\right)\right), \tag{12}$$

where $C$ is the objective function, generally called the correlation coefficient in DIC. The Newton-

Raphson method [17] is used to find the solution of the optimization problem above using an iteration scheme as

$$\nabla\nabla C\left(\mathbf{a}^{(l)}\right)\left(\mathbf{a}^{(l+1)}-\mathbf{a}^{(l)}\right)=-\nabla C\left(\mathbf{a}^{(l)}\right)\qquad l=0,1,2,\cdots \tag{13}$$

where $\mathbf{a}^{(l)}$ represents the solution of **a** from the iteration in the $l$-th step.

Given a specific expression of the correlation coefficient, the solving scheme can be constructed. For the simplest definition of the correlation coefficient,

$$C = \sum_{\Omega_j} \left[ f(\mathbf{X}) - G_i^{(m)}\left(\mathbf{X} + \mathbf{U}(\mathbf{X}, t; \mathbf{a})\right) \right]^2, \tag{14}$$

the first-order derivative is calculated as

$$\nabla C = \frac{\partial C}{\partial \mathbf{a}} = -2\sum_{\Omega_j} (f - G) \frac{\partial G}{\partial \mathbf{a}}, \tag{15}$$

where $f$ is short for $f(\mathbf{X})$, $G$ is short for $G_i^{(m)}\left(\mathbf{X} + \mathbf{U}(\mathbf{X}, t; \mathbf{a})\right)$, and

$$\frac{\partial G}{\partial \mathbf{a}} = \frac{\partial G}{\partial \tilde{x}}\frac{\partial \tilde{x}}{\partial \mathbf{a}} + \frac{\partial G}{\partial \tilde{y}}\frac{\partial \tilde{y}}{\partial \mathbf{a}} = \frac{1}{2m+1}\int_{t_{i-m}}^{t_{i+m}} \begin{bmatrix} 1 & 0 \\ 0 & 1 \\ \Delta x & 0 \\ 0 & \Delta x \\ \Delta y & 0 \\ 0 & \Delta y \\ \Delta t & 0 \\ 0 & \Delta t \end{bmatrix} \begin{bmatrix} \frac{\partial g_t}{\partial \tilde{x}} \\ \frac{\partial g_t}{\partial \tilde{y}} \end{bmatrix} dt, \tag{16}$$

where $\tilde{x} = x + u, \tilde{y} = y + v$ are the new coordinates of the point.

The second-order derivative, usually called the Hessian matrix, is calculated as

$$\nabla\nabla C = \frac{\partial^2 C}{\partial \mathbf{a}\partial \mathbf{a}} = 2\sum_{\Omega_j} \left[ \frac{\partial G}{\partial \mathbf{a}}\frac{\partial G}{\partial \mathbf{a}} - \frac{\partial^2 G}{\partial \mathbf{a}\partial \mathbf{a}}(f - G) \right]. \tag{17}$$

If **a** is closed to the solution, $f \approx G$, Eq. 17 can be simplified [18] as

$$\nabla\nabla C = 2\sum_{\Omega_j} \left( \frac{\partial G}{\partial \mathbf{a}}\frac{\partial G}{\partial \mathbf{a}} \right). \tag{18}$$

Substituting Eq. 16 into Eqs. 15 and 18, the first- and second-order derivatives of $C$ are obtained. Then, substituting these into Eq. 13, the iteration can be realized by giving an appropriate initial solution. With the iteration, the new DIC scheme is solved.

## Verification and Application of STS-DIC

The proposed STS-DIC is verified in this section using speckle image series from numerical simulation and real experiments.

### *Verification using a simulated image series*

The simulated image series consists of ten images: one reference image and nine deformed images. The reference image, with size $301 \times 301$ pixels, was cropped from a real speckle image as shown in

Fig. 5. The deformed images were simulated to take the displacement field of different rigid translations using the simulation algorithm given in [10]. In total, nine deformed images were simulated with the horizontal displacement $u$ expressed as

$$u(\mathbf{X},t)=\frac{t}{20} \qquad (t = 1, 2, 3\ldots9). \tag{19}$$

Once all the images had been created, Gaussian noise with zero mean and standard variation of $\sigma_g$ (gray level with respect to an 8-bit image) was added to all the images, that is, both the reference and deformed images. To investigate speckle images with different degrees of noise, 11 levels of noise, i.e., $\sigma_g$ =0, 1, 2… 10, were included and 11 speckle image series were created.

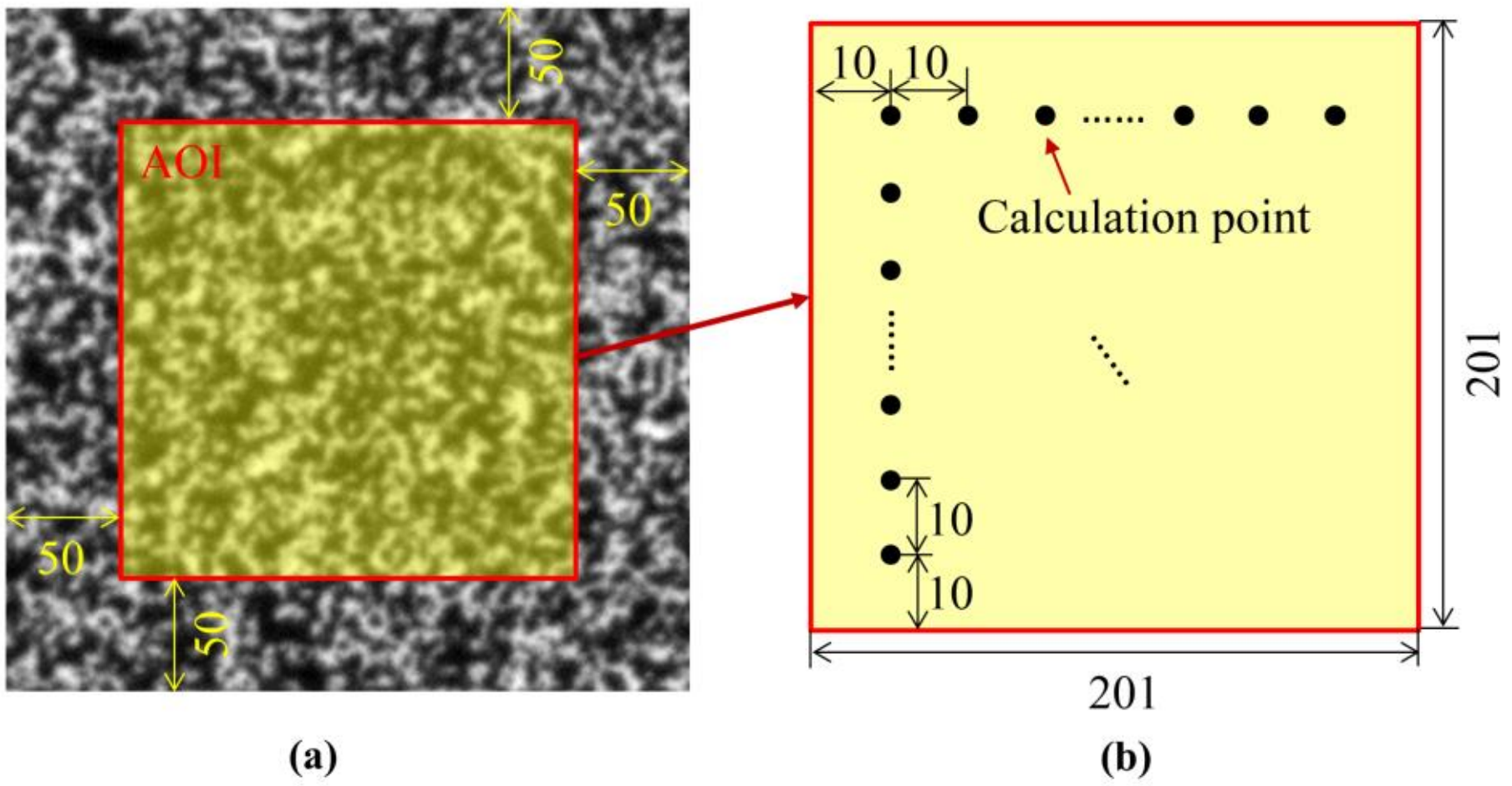

**Fig. 5** Reference speckle image: AOI and the distribution of calculation points

For each series, the displacement of 19×19 points shown in Fig. 5b was calculated using STS-DIC with a subset size of 21×21 pixels. The horizontal displacement ($u$) of the 361 points was quantitatively compared with the given displacement, and the error evaluated as

$$e=\sqrt{\frac{1}{361}\sum_{k=1}^{361}\left(u_k-\tilde{u}_k\right)^2}\,, \tag{20}$$

where $u_k, \tilde{u}_k$ are the given and measured displacements of the $k$-th point, respectively.

The effect of STS-DIC is first expressed by comparing the results from two of the image series with and without noise, i.e., $\sigma_g$ = 0 and $\sigma_g$ = 5. The errors of STS-DIC with different numbers of superposed deformed images (2$m$+1) are shown in Table 1.

**Table 1** Error ($e$) of the results from traditional DIC and STS-DIC

| Number of averaged images | $e$ ($\times10^{-3}$ gray level) | | time (s) |
|---|---|---|---|
| | $\sigma_g$ =5 | $\sigma_g$ =0 | |
| 1 (traditional) | 10.7272 | 1.1779 | 2.9742 |
| 3 | 5.5453 | 1.6364 | 53.2310 |
| 5 | 4.7422 | 1.6434 | 95.7100 |
| 7 | 4.4291 | 1.8627 | 120.7580 |
| 9 | 3.7261 | 1.7438 | 143.8548 |

* The PC used for the calculations was an Intel(R) Core (TM) i5-4200U 4-core CPU (1.60 GHz) with 8-GB memory.

It can be seen from the table that STS-DIC outperforms traditional DIC when processing images with noise ($\sigma_g$ = 5), because of the improvement in quality achieved by averaging the speckle images. For the simple deformation (translation) in this example, the error of STS-DIC dramatically decreases compared with that of traditional DIC. Furthermore, with an increase in the number of images averaged, the error of STS-DIC decreases. However, for images without noise ($\sigma_g$ = 0), STS-DIC performs worse than traditional DIC. This is understandable because STS-DIC processes more images and performs an optimization with more unknowns; in this case, the superposing does not improve the quality of the image. The time consumption for DIC and STS-DIC was also recorded and is shown in the table. It is reasonable that STS-DIC is much more time consuming than traditional DIC. Moreover, time consumption increases with an increasing number of images averaged.

The results of all 11 image series are shown in Fig. 6. For each series, the results are given for three different situations, i.e., traditional DIC, STS-DIC (with five images averaged), and the direct smoothing of the traditional DIC results. For direct smoothing, the DIC results were first calculated and then smoothed over time using the same time span, i.e., five images.

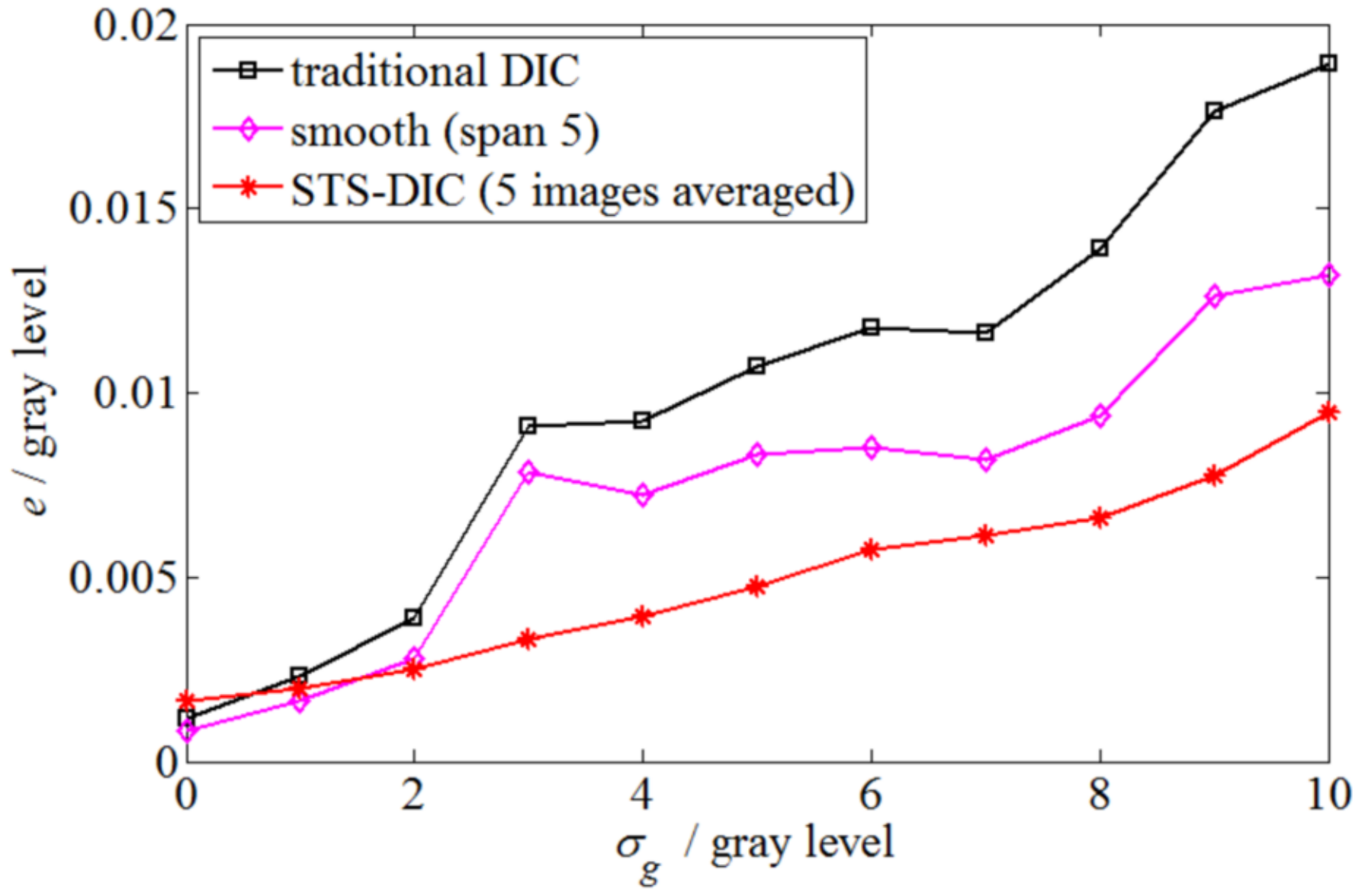

**Fig. 6** Error of STS-DIC when processing speckle images with different noise levels

Results shown in Fig. 6 correlate with those in Table 1. For image series with heavy noise ($\sigma_g \geq 2$), STS-DIC outperforms traditional DIC. With increasing noise, the improvement also increases. Directly smoothing the DIC results helps to improve the calculation for this simple deformation, but the accuracy is still much worse than that of STS-DIC. For image series with minimal noise ($\sigma_g$ = 0 and 1 in this case), STS-DIC does not show any accuracy improvement. This has been explained above.

*Verification using speckle images from a real experiment: uniaxial tension*

A polycarbonate specimen, shown in Fig. 7a, was tensioned uniaxially and the strain in two directions, i.e., along and perpendicular to the loading direction, were measured using DIC. The WDW-50E test machine was used to tension the specimen with a loading speed of 0.2 mm/min. The speckle images were recorded with a Basler A641f digital CCD camera, with a resolution of 1624×1236 pixels and frame rate of 1 frame per second. After the experiment, the deformation within the AOI (shown in Fig. 7b) was obtained using traditional DIC and STS-DIC. For comparison, two strain gauges were bonded on the opposite surface to the speckle and centered on the AOI. The validity and advantage of the proposed method were then investigated by comparing the results of STS-DIC with those of the strain gauges and traditional DIC.

In total, 11×5 points within the AOI were processed with traditional DIC and STS-DIC (with 5 images averaged) to obtain the displacement fields. The subset size in the DIC calculation was chosen as 31×31 pixels. After the displacement calculation, the normal strain in the two directions was calculated by fitting the displacement field as a plane. Images from 60 different loading levels within the elastic stage were chosen for error analysis. To compare the DIC results, results of the strain gauges were taken as standard, with the error evaluated as

$$\begin{aligned} e_x &= \sqrt{\frac{1}{60}\sum\left(\varepsilon_x^{\mathrm{D}} - \varepsilon_x^{\mathrm{S}}\right)^2} \\ e_y &= \sqrt{\frac{1}{60}\sum\left(\varepsilon_y^{\mathrm{D}} - \varepsilon_y^{\mathrm{S}}\right)^2} \end{aligned}, \tag{21}$$

where $\varepsilon_x^{\mathrm{D}}$ and $\varepsilon_y^{\mathrm{D}}$ denote the strain in the *x*- and *y*-directions measured by DIC, and $\varepsilon_x^{\mathrm{S}}$ and $\varepsilon_y^{\mathrm{S}}$ denote the strain in the *x*- and *y*-directions measured by the strain gauge, respectively.

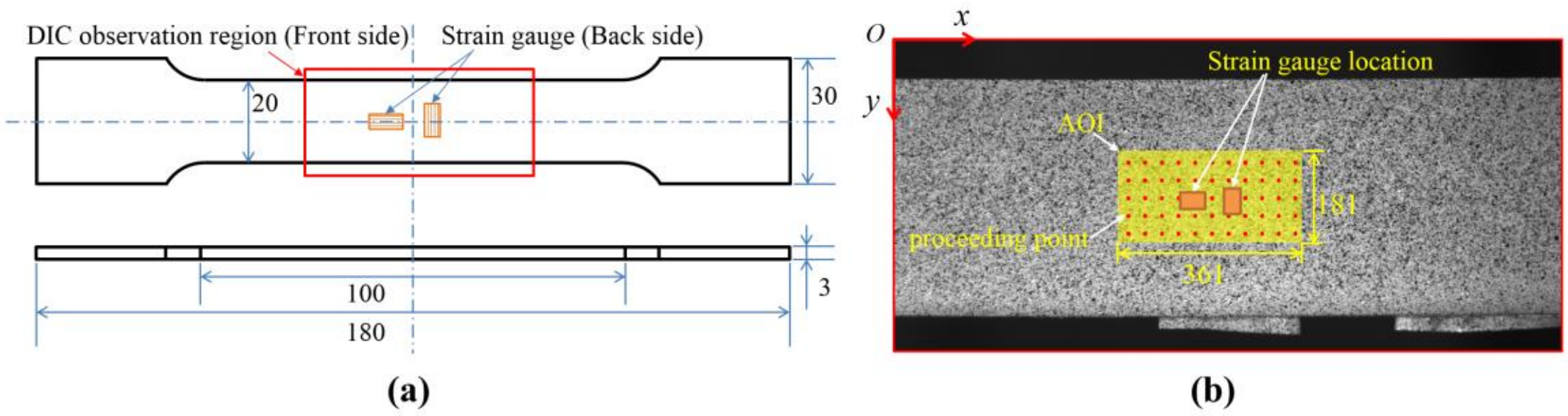

**Fig. 7** Experimental setup for the uniaxial tension of polycarbonate: **(a)** dimensions of the specimen (in mm), and **(b)** speckle image and AOI (in pixels)

**Table 2** Error between the results from DIC (traditional DIC and STS-DIC) and the strain gauge

| | traditional DIC | STS-DIC (5 images averaged) |
|---|---|---|
| $e_x$ | 58.4620 | 18.8731 |
| $e_y$ | 116.5760 | 57.3353 |

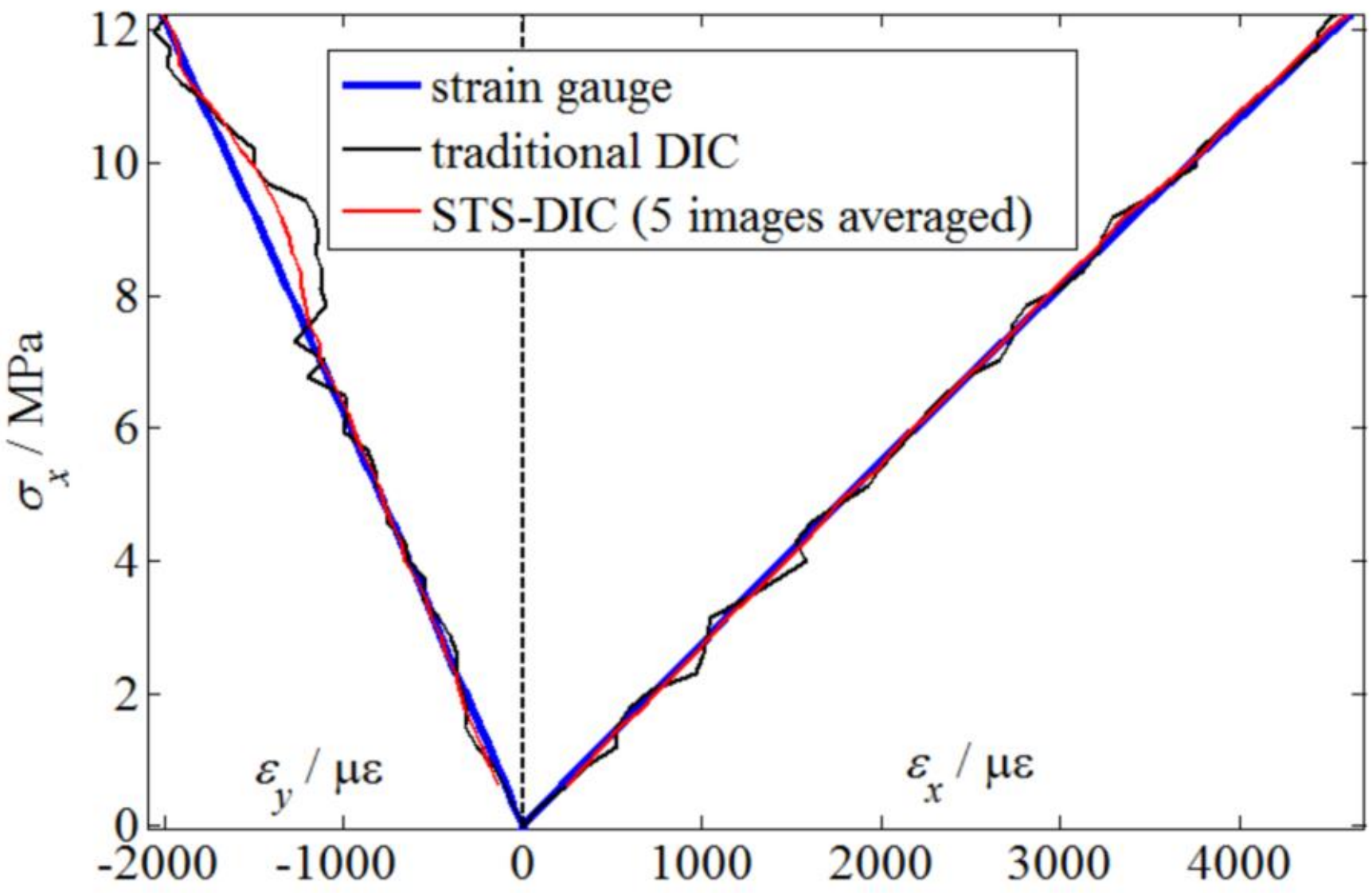

**Fig. 8** Stress-strain curves measured by STS-DIC, DIC, and their deviation from that of the strain gauge

The errors of the DIC results compared with those of the strain gauge are shown in Table 2. It is seen that STS-DIC decreases the error with an amplitude of 1/2 (in the *y*-direction) to 2/3 (in the *x*-direction). The stress-strain curves measured by STS-DIC, DIC, and the strain gauge are shown in Fig. 8. It can be seen that the curve measured by STS-DIC is smoother and closer to the standard curve measured by the strain gauge, compared with the result obtained by traditional DIC. The results in Table 2 and Fig. 8 show that STS-DIC improves the measurement accuracy of traditional DIC in this practical tension experiment.

*Verification using speckle images from a real experiment: defect detection*

The two previous examples show the validity and advantage of STS-DIC in the case of simple deformation, i.e., translation and homogeneous deformation. In this section, STS-DIC is also verified with more complicated deformation, i.e., deformation induced by a defect on a tensioned specimen.

As shown in Fig. 9a, an impenetrable hole is manufactured on a polycarbonate tensile specimen to simulate a defect. A speckle pattern is painted on the surface opposite the hole and DIC is used to measure the deformation field. Although the defect is not seen on the surface, it will induce a distortion to the original deformation pattern (as shown in Fig. 10a), from which it can be detected. The specimen was loaded with the same test machine and same loading speed as in the previous experiment, while the images were captured by a camera with a higher resolution, i.e., an IPX-16M3-L CCD camera with a resolution of 4872 ×3248 pixels. After the experiment, the displacement and strain field within the AOI (as show in Fig. 9b) were calculated using DIC, with a subset size of 31 ×31 pixels. For STS-DIC, five images were averaged.

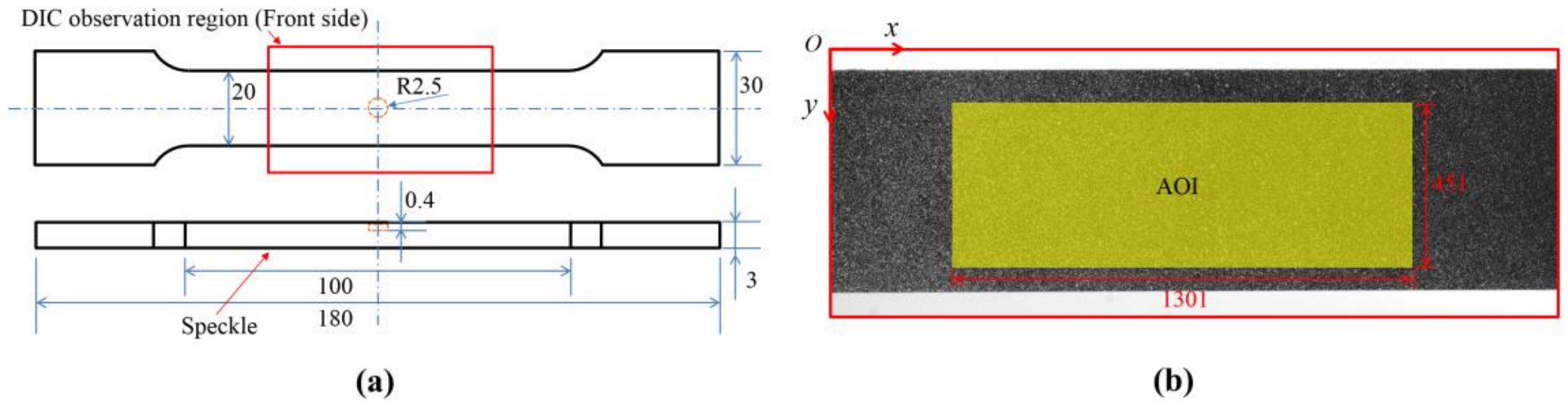

**Fig. 9** Experimental setup for defect detection: **(a)** dimensions of the PC specimen with a defect (in mm), and **(b)** the speckle image and AOI (in pixels)

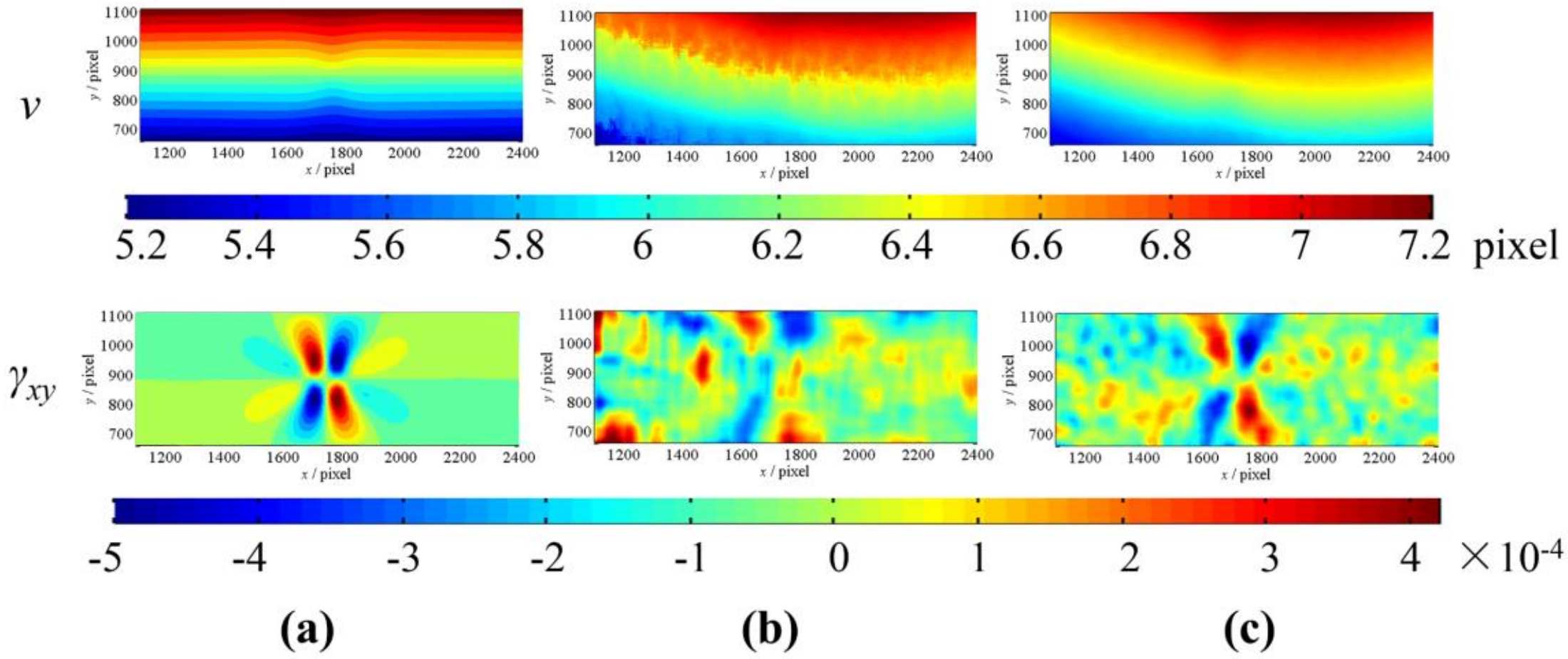

**Fig. 10** Displacement and strain fields: **(a)** simulated result, **(b)** traditional DIC result, and **(c)** STS-DIC result (in pixels)

The vertical (perpendicular to the tension direction) displacement field and the shear strain field calculated by DIC are shown in Fig. 10b (traditional DIC) and in Fig. 10c (STS-DIC). This figure shows that the displacement and strain fields measured by STS-DIC are much smoother and more consistent with the theoretical results than those of traditional DIC. This indicates that STS-DIC can improve the measurement accuracy of traditional DIC in a complicated deformation situation.

## Conclusion

We incorporated temporal continuity of deformation in DIC and developed STS-DIC. In STS-DIC, several adjacent deformed speckle images are treated as a whole for comparison with the reference image. On the contrary, only a single deformed speckle image is compared with the reference image in traditional DIC. The displacement of several adjacent images over time is assumed to be continuous and can be expressed as a simple linear relation. The introduction of temporal continuity in STS-DIC results in an improvement in the accuracy of DIC, in the same way as the introduction of spatial continuity improves global DIC.

The STS-DIC scheme was constructed by deriving expressions based on the basic assumption of DIC analysis, i.e., the “constant intensity assumption” between the reference and deformed images. An optimization model and optimization scheme were created. Speckle images from both simulations and experiments were used to verify the proposed STS-DIC. The results show that STS-DIC greatly improves the accuracy when measuring both simple deformation (translation and homogeneous tension) and complicated deformation (strain concentration from a defect).

More interestingly, the results of STS-DIC were also compared with the direct smoothing of the DIC results, showing that the former method outperforms the latter. This means that inherent smoothing, i.e., “smoothing” of deformed images, is more effective than “external” smoothing of the result data, thereby showing the importance of the modification of the model.

The assumption of STS-DIC, i.e., linear deformation of several images over time, can easily be satisfied in most circumstances, with either static or dynamic loading. Even for some special cases, for example, the occurrence of a crack, STS-DIC can still be used in a series that excludes those images when the crack has just appeared, and also in areas with no crack. Nevertheless, the series in STS-DIC should be chosen carefully in a practical measurement, especially when the deformation is complicated. Too long series greatly increases the computational cost without improving accuracy.

In this paper, STS-DIC was developed based on a subset spatial approach with consideration of time cost. However, the derivation in the second section also suggests the possibility of STS-DIC based on a global spatial approach.

**Acknowledgments**

The authors would like to thank the support from the National Science Foundation of China (11172039, 11372038).